\documentclass[prb,aps,twocolumn,showpacs,superscriptaddress]{revtex4}
\usepackage{graphicx}
\usepackage{wasysym}

\newcommand{\beq}{\begin{equation}}
\newcommand{\eeq}{\end{equation}}
\newcommand{\bea}{\begin{eqnarray}}
\newcommand{\eea}{\end{eqnarray}}

\begin{document}
	
\title{Quantum order by disorder in a spin-one frustrated magnet on the kagome lattice}

%\author{S. V. I.}

\author{Sergei V. Isakov}
\affiliation{ 
Institute for Theoretical Physics, ETH Z\"urich, CH-8093 Z\"urich, Switzerland} 
\author{Yong Baek Kim}
\affiliation{Department of Physics, University of Toronto, Toronto,
Ontario M5S 1A7, Canada}

\date{\today}

\begin{abstract}
We study the XXZ spin-one quantum magnet on the kagome lattice as an example 
where quantum fluctuations on highly degenerate classical ground states lead to
various exotic quantum ground states. Previous studies have predicted several 
quantum phases, but different analytical approaches do not necessarily lead to the
same physical picture. In this work, we use Quantum Monte Carlo computations
to critically examine some of the predictions made in the string-net mean-field theory
and the degenerate perturbation theory combined with duality analysis 
and effective field theory. It is found that the resulting phase diagram differs from
some of the previous predictions. Further implications of our results to different analytical
approaches are discussed.
\end{abstract}


\maketitle

\section{Introduction}

Macroscopic degeneracy of classical ground states in frustrated magnets is a fertile 
ground for emergence of unusual quantum ground states such as 
spiral magnetic order, quantum spin liquid, and valence bond solid (VBS)
that occur via quantum fluctuations. \cite{sachdev}
Any perturbation on the degenerate classical ground states, however, is inherently
strong and there may also be more than one competing ground states that are extremely 
close in energy. The situation is very reminiscent of quantum Hall states that emerge from
highly degenerate Landau levels. This is indeed one of the reasons 
why identification of true quantum ground state in frustrated magnets is such a difficult task.

%---------------------------------------------------------------------------
\begin{figure}[t]
\includegraphics[width=3.0in]{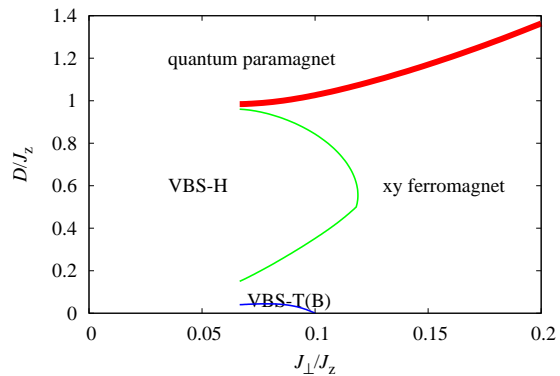}
\caption{ (color online).
The schematic phase diagram. First order phase transitions are denoted by
thin lines (blue and green online) and a continuous phase transition is
denoted by a thick line (red online).
}
\label{fig:phase:diagram}
\end{figure}
%---------------------------------------------------------------------------

Different kinds of non-perturbative analytical approaches have been proposed in literature to
understand this quantum order (or disorder) by disorder phenomena. These approaches
include various degenerate perturbation theories, \cite{balents}
duality analysis combined with effective field
theories, \cite{sachdev,balents}
and string net condensation picture. \cite{wen}
Since different approaches may not necessarily
lead to the same conclusion, it is important to understand the limitations of various approaches.
One useful way to obtain such information is doing unbiased numerics by appropriately 
choosing concrete examples where predictions from different approaches may vary.

Here we consider the following spin-one XXZ model on the kagome lattice as 
such an example.
\beq
  H=-J_\perp\sum_{\langle ij \rangle} (S^x_iS^x_j+S^y_iS^y_j)
    + J_z \sum_{\langle ij \rangle} S^z_i S^z_j + D \sum_i (S^z_i)^2,
  \label{model}
\eeq
where $J_\perp, J_z, D > 0$, the first two sums run over the nearest neighbors on the kagome
lattice, and $D$ is the strength of the single-ion anisotropy. 
This model is also equivalent to a boson model with nearest neighbor repulsive interaction
where boson occupation number can only assume $0, 1, 2$. 
Previously two different analytical approaches have been used to study the phase
diagram of this model. Using degenerate perturbation theory and mapping to
a dimer model combined with duality analysis, Xu and Moore \cite{xu:moore1,xu:moore2}
predict that there
exist three different phases: an XY ferromagnetic phase for $J_\perp \gg J_z, D$,
a plaquette valence bond solid (VBS-H) phase with resonating hexagons for
$J_\perp \ll D < J_z$, and a gapped ``photon" (quantum paramagnetic) phase for 
$J_\perp \ll J_z < D$. 
The gapped ``photon" phase is a descendant of an unstable two-dimensional 
quantum spin liquid phase with linearly dispersing neutral ``photon" modes.
It was also predicted that there would be a direct
transition between VBS-H and a gapped ``photon" phase.
Using a quite different approach, Levin and Wen \cite{levin:wen} proposed a 
mean-field theory where a class of variational wavefunctions
based on the so-called ``string-net" picture was used to map out the global 
phase diagram. While they also predict the existence of a plaquette
phase as well as a gapped ``photon" phase, their plaquette phase is
characterized by frozen spin configurations and this frozen plaquette (FP) phase 
is different from VBS-H in lattice and spin symmetries. \cite{levin:wen}
In this string-net mean-field theory, the ``photon" phase corresponds to 
the string-net condensed phase \cite{levin:wen} and the unstable two-dimensional
spin liquid phase mentioned above. This ``photon" phase 
was found in a narrow region around $J_z\sim D$ if one
ignored the nonperturbative instanton contribution. The ``photon" excitation
acquires a finite gap, however, due to the instanton effect 
in 2+1 dimensions. \cite{levin:wen,polyakov} The resulting state is
an ordinary quantum paramagnetic state.

In this work, we map out the phase diagram of the model given by
Eq.~\ref{model} using quantum Monte Carlo method. We use a plaquette
generalization \cite{sse:improved} of the stochastic series expansion
algorithm. \cite{sse} We consider only the parameter region $J_\perp>0$,
$J_z>0$, and $D>0$. The schematic phase diagram is shown in
Fig.~\ref{fig:phase:diagram}. First, we find that a plaquette ordered VBS
phase indeed arises and it is the VBS-H phase, not the FP.
In addition to the phases discussed in the previous works, we also find 
an additional phase with resonating lattice units for small values of
$D/J_z$ and $J_\perp/J_z$. The ground state in this region is either a
phase with resonating triangles (VBS-T) or a phase with resonating bow
ties (VBS-B).
It was argued in
Ref.~\onlinecite{xu:moore2} that there should be a direct continuous 
transition from the VBS-H phase to the paramagnetic (gapped ``photon'') phase. 
We do not observe such a transition for $J_\perp/J_z \gtrsim 0.08$. However,
we cannot fully rule out the possibility of a direct VBS-H-paramagnet
phase transition at even smaller values of $J_\perp/J_z$.

The rest of the paper is organized as follows. In Section II, we will
discuss the details of the phase diagram and properties of all the
phases discovered in the numerics. Here we also discuss the
difference between our results and the predictions of previous works.
We discuss the implications of our results to various analytical approaches 
in Section III.

%---------------------------------------------------------------------------
\section{Construction of the phase diagram}

%---------------------------------------------------------------------------
\subsection{VBS-H phase}

%---------------------------------------------------------------------------
\begin{figure}[t]
\includegraphics[width=3.0in]{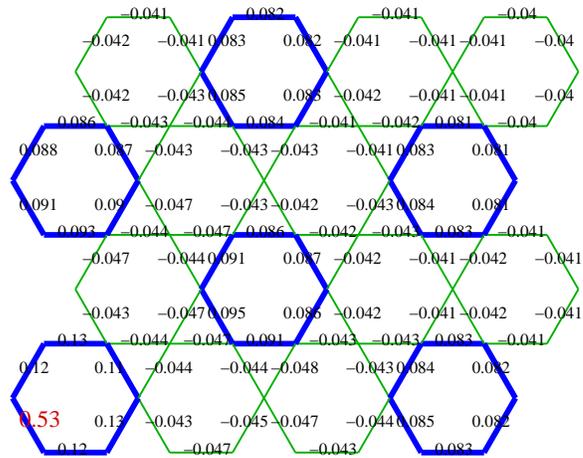}
\caption{ (color online).
The correlation function $C_{\text b}({\mathbf r}_1-{\mathbf r}_\delta)$
between the bond indicated by a large text (red online) and the other bonds
for $L=24$, $J_z/J_\perp=10$, $D/J_z=0.4$, and $T=J_\perp/48$. Resonating
hexagons are denoted by thick lines.
}
\label{fig:bchi:h}
\end{figure}
%---------------------------------------------------------------------------

%---------------------------------------------------------------------------
\begin{figure}[t]
\includegraphics[width=3.0in]{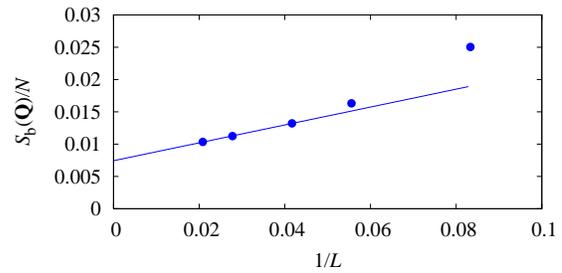}
\caption{ (color online).
Finite size scaling of the equal time bond-bond structure factor
$S_{\text b}(\mathbf{Q}_0)$ for $L=24$, $J_z/J_\perp=10$, $D/J_z=0.4$,
and $T=J_\perp/48$. In this and the other structure factor plots, error
bars (if not visible) are smaller than the symbol sizes, and the line
shows a linear extrapolation to the thermodynamic limit.
}
\label{fig:bsf:h}
\end{figure}
%---------------------------------------------------------------------------

%---------------------------------------------------------------------------
\begin{figure}[t]
\includegraphics[width=3.0in]{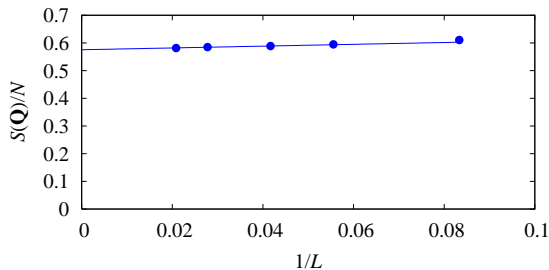}
\caption{ (color online).
Finite size scaling of the equal time spin-spin structure factor
$S({\mathbf Q}_0)$ for $L=24$, $J_z/J_\perp=10$, $D/J_z=0.4$,
and $T=J_\perp/48$.
}
\label{fig:sf:h}
\end{figure}
%---------------------------------------------------------------------------

%---------------------------------------------------------------------------
\begin{figure}[t]
\includegraphics[width=3.0in]{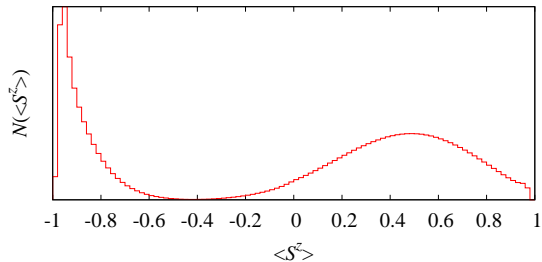}
\caption{ (color online).
Probability distribution (arbitrary units) of $\bar{S}^z_i$ for
$L=48$, $J_z/J_\perp=10$, $D/J_z=0.4$, and $T=J_\perp/48$.
The distribution function is independent of the system size (not shown).
}
\label{fig:sz:dist:h}
\end{figure}
%---------------------------------------------------------------------------

We begin with the analysis of the VBS-H phase that is found in a
wide lobe as shown in Fig.~\ref{fig:phase:diagram}. This phase has plaquette
order. Two different plaquette phases were proposed previously for this
region of parameters. In the plaquette phase (VBS-H) predicted by Xu and Moore,
\cite{xu:moore1,xu:moore2} spins resonate on a third of the total number of
hexagons. These spins have $\langle S^z_i \rangle=b/2$ around 
resonating hexagons, where $b$ is some number. 
The rest of the spins is fixed to $\langle S^z_i \rangle=-b$. In the
frozen plaquette phase (FP) predicted by Levin and Wen, \cite{levin:wen}
spins are frozen on a third of all the hexagons with
alternating values of $\langle S^z_i \rangle=\pm a$ around those frozen
hexagons; the rest of the spins
has $\langle S^z_i \rangle=0$. \cite{levin:wen} Thus, these two phases break Ising
and lattice symmetries in different ways.
Note that these two phases can also be distinguished
by the distribution function of $\langle S^z_i \rangle$ 
(more discussions are given below).

In the resonating plaquette phase, the bond-bond correlation functions
should have peaks in momentum space corresponding to the specific
pattern in real space characterizing plaquette bond order.
In Fig.~\ref{fig:bchi:h}, we show the bond-bond correlations function in
real space that is given by
\beq
  C_{\text b}({\mathbf r}_\gamma-{\mathbf r}_\delta)=\left\langle
    \left[\frac{1}{\beta}\int B_{\gamma\tau}d\tau-B_0 \right]
    \left[ \frac{1}{\beta}\int B_{\delta\tau}d\tau-B_0 \right] \right\rangle,
\label{eq:bchi}
\eeq
where $B_{\alpha(i,j), \tau}=J_\perp(S^x_iS^x_j+S^y_iS^y_j)$ is the
off-diagonal bond operator (at imaginary time $\tau$) of the bond $\alpha$
connecting spins $i$ and $j$, and $B_0$ denotes the background bond strength.
In Fig.~\ref{fig:bchi:h}, one can clearly see the pattern that is compatible
with the VBS-H phase.
To confirm that the bond order is long-ranged, we show in
Fig.~\ref{fig:bsf:h} the finite size scaling of the equal time bond-bond
structure factor $S_{\text b} ({\mathbf q})$ 
at the ordering wavevector $\mathbf{q}=\mathbf{Q}_0=(4\pi/3,0)$, 
\beq	
  S_{\text b}({\mathbf q})=N\langle B^{\dagger}_{{\mathbf q}\tau}
    B_{{\mathbf q}\tau} \rangle,
\label{eq:bsf}
\eeq
where $B_{{\mathbf q}\tau}=(1/N)\sum_\alpha B_{\alpha \tau}
\exp(i{\mathbf q}{\mathbf r_\alpha})$. Note that the expression of 
$B_{{\mathbf q}\tau}$ involves the sum over the bond index
$\alpha$ and $N$ is the number of sites. 
The structure factor divided by the number of sites clearly goes to a finite 
value in the thermodynamic limit
even though the extrapolated value is quite small.

On the other hand, the bond-bond correlation described in the previous paragraph 
might also be compatible with the FP phase.
More robust diagnostic is the distribution function
of local $\langle S^z_i\rangle$, which should have two peaks for the VBS-H
phase and three peaks for the FP phase if the system is in one of the
six degenerate symmetry-broken states. \cite{remark}
For each Monte Carlo configuration,
we compute the time-averaged $\bar{S}^z_i=(1/\beta)\int_0^\beta d\tau S^z_{i\tau}$,
$S^z_{i\tau}$ is the $z$-component of the spin operator at site $i$ and
imaginary time $\tau$. 
In Fig.~\ref{fig:sz:dist:h}, the distribution function of
$\bar{S}^z_i$ is shown. The distribution function has two peaks;
one sharp peak at $-b$ and the other broader peak at $b/2$. This is what
we expect for the VBS-H phase. For the frozen plaquette phase, one expects
three peaks at $a$, $0$, and $-a$.

The VBS-H phase also has magnetic order. In Fig.~\ref{fig:sf:h}, we show
the finite size scaling of the equal time spin-spin structure factor at
$\mathbf{q}=\mathbf{Q}_0$,
\beq
  S({\mathbf q})=N\langle S^{\dagger}_{{\mathbf q}\tau}
    S_{{\mathbf q}\tau} \rangle,
\label{eq:sf}
\eeq
where
$S_{{\mathbf q}\tau}=(1/N)\sum_i S^z_{i\tau}\exp(i{\mathbf q}{\mathbf r_i})$.

Therefore, our Monte Carlo data are consistent with the resonating plaquette phase (VBS-H). 
It is worth noting that the VBS-H phase is quite similar to the VBS phase
discovered in the hard-core boson model on the kagome lattice at fillings
of $1/3$ and $2/3$ (for details, see Ref.~\onlinecite{kagome1,kagome2}).

%---------------------------------------------------------------------------
\subsection{VBS-T(B) phase}

%---------------------------------------------------------------------------
\begin{figure}[t]
\includegraphics[width=3.0in]{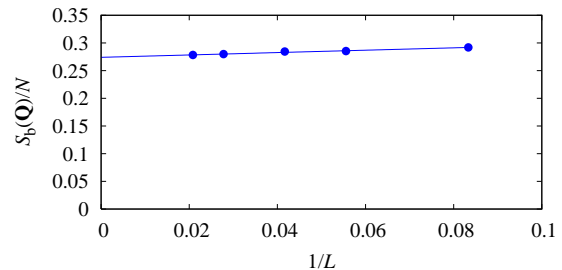}
\caption{ (color online).
Finite size scaling of the equal time bond-bond structure factor
$S_{\text b}(\mathbf{Q}_3)$ for $J_z/J_\perp=10.5$, $D=0$, and $T=J_\perp/12$. 
}
\label{fig:bsf:d0}
\end{figure}
%---------------------------------------------------------------------------

%---------------------------------------------------------------------------
\begin{figure}[t]
\includegraphics[width=3.0in]{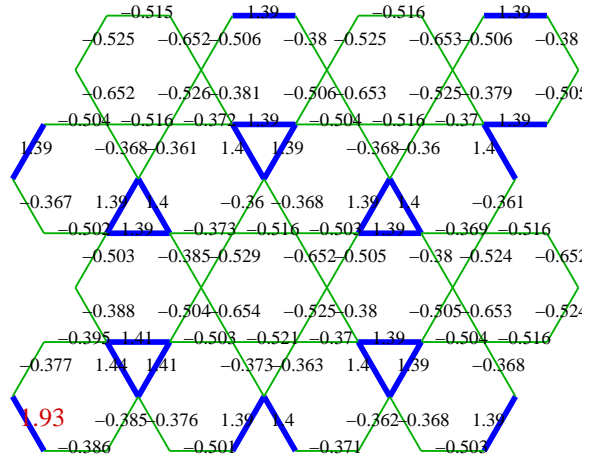}
\caption{ (color online).
The correlation function $C_{\text b}({\mathbf r}_1-{\mathbf r}_\delta)$
between the bond indicated by a large text (red online) and the other bonds
for $L=12$, $J_z/J_\perp=10.5$, $D=0$, and $T=0.05J_\perp/$. Resonating
triangles (trimers) are denoted by thick lines.
}
\label{fig:bchi:d0}
\end{figure}
%---------------------------------------------------------------------------

%---------------------------------------------------------------------------
\begin{figure}[t]
\includegraphics[width=3.0in]{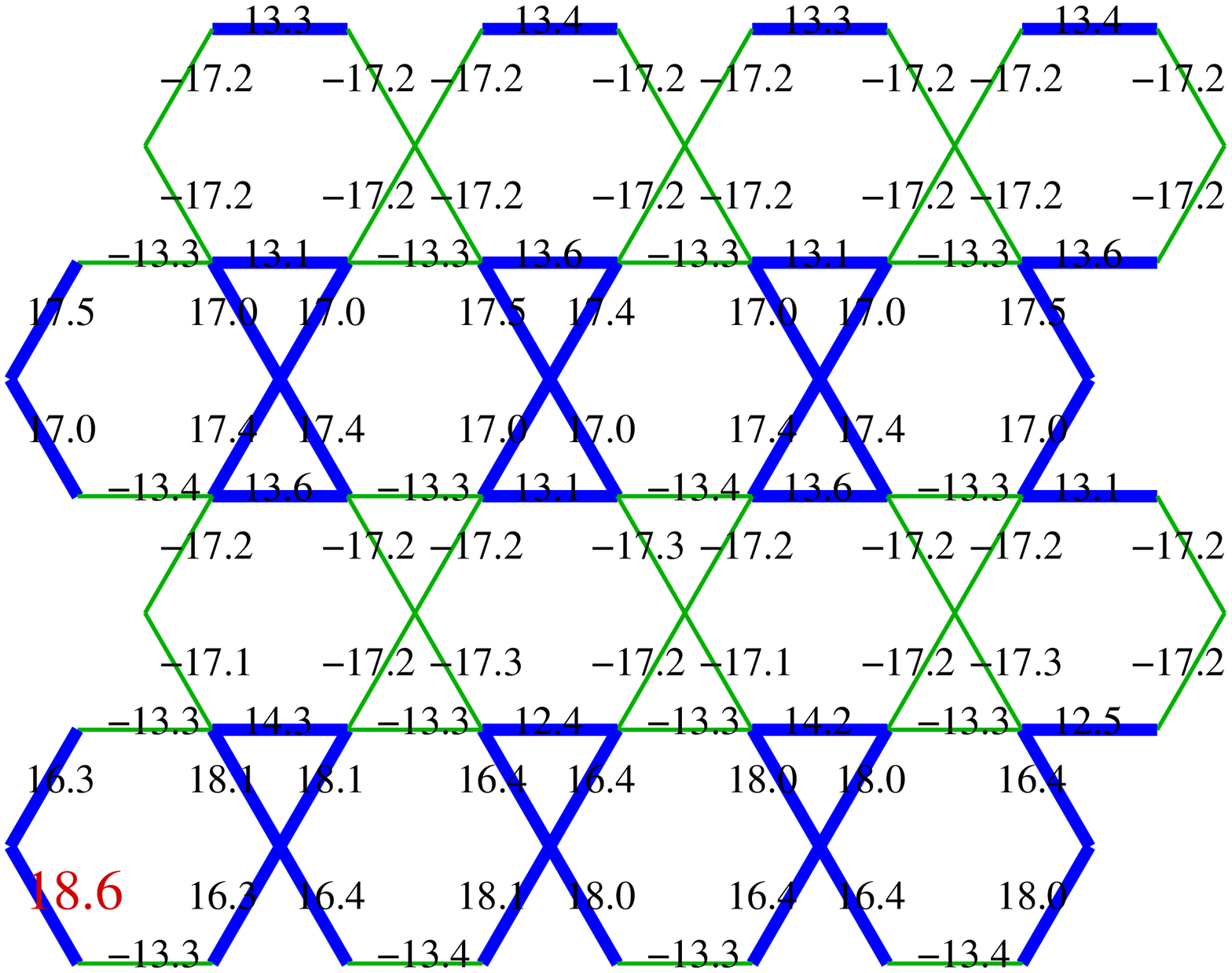}
\caption{ (color online).
The correlation function $C_{\text b}({\mathbf r}_1-{\mathbf r}_\delta)$
between the bond indicated by a large text (red online) and the other bonds
for $L=12$, $J_z/J_\perp=10.5$, $D=0$, and $T=0.002J_\perp$. Resonating
bow ties are denoted by thick lines.
}
\label{fig:bchi:d0b}
\end{figure}
%---------------------------------------------------------------------------

%---------------------------------------------------------------------------
%\begin{figure}[t]
%\includegraphics[width=3.0in]{sf_jz=10.5_d=0_b=12.eps}
%\caption{ (color online).
%Finite size scaling of the equal time spin-spin structure factor
%$S({\mathbf Q}_3)$ for $L=24$, $J_z/J_\perp=10.5$, $D=0$, and $T=J_\perp/12$.
%}
%\label{fig:sf:d0}
%\end{figure}
%---------------------------------------------------------------------------

Another interesting result in this work is the discovery of another
phase with resonating plaquettes at small values of $D/J_\perp$ and large
values of $J_z/J_\perp$ as shown in Fig.~\ref{fig:phase:diagram}. This phase
was not predicted before. There are actually two different competing phases
that are very close in energy. Triangles resonate in one of those phases
(VBS-T phase) and bow-ties resonate in the other one (VBS-B phase). We are
not able to determine reliably which phase is the true ground state.

We find that there is a finite temperature transition from the high
temperature paramagnetic phase to the resonating phase for $J_z/J_\perp\ge 10$
and $D$ close to zero. We have not attempted to obtain the precise location
of this transition. The transition temperature $T_0\approx J_\perp/10$ for
$J_z/J_\perp=10.5$ and $D=0$.
Typically, we find the VBS-T phase just below the transition
and either the VBS-T or VBS-B phase at much lower temperatures
depending on the configuration we start our Monte Carlo simulations with.

The bond-bond correlators have well pronounced peaks at
$\mathbf{Q}_1=(\pi,0)$, $\mathbf{Q}_2=(0,\pi)$, and $\mathbf{Q}_3=(\pi,\pi)$
in the VBS-T phase and at those or other points in the VBS-B phase.
In Fig.~\ref{fig:bsf:d0}, the finite size scaling of the equal time bond-bond
structure factor $S_{\text b}(\mathbf{Q}_3)$ defined in  Eq.~\ref{eq:bsf}
is shown for the VBS-T phase. The structure factor divided by the
number of sites scales to a finite value in the thermodynamic limit
indicating long range valence bond order.
Note that the VBS order parameter is very large \cite{beach:sandvik} in 
 sharp contrast
to the VBS-H bond order parameter and to some other models, in which the
VBS order was confirmed in quantum Monte Carlo simulations. \cite{kagome1}
To investigate the nature of the VBS-T(B) phase, we study the real space
correlation function defined in Eq.~\ref{eq:bchi}. As shown in
Figs.~\ref{fig:bchi:d0} and ~\ref{fig:bchi:d0b}, the VBS-T phase exhibits
a network of resonating triangles and the VBS-B phase shows a network of
resonating bow-ties.

It is worth noting that the presence of the insulating phase at $D=0$ is in
sharp contrast to the spin-$1/2$ XXZ model with ferromagnetic XY and
antiferromagnetic Ising exchange
interactions on the kagome lattice, where the uniform XY ferromagnet
persists at any finite value of $J_\perp/J_z$. \cite{kagome1, kagome2}

%---------------------------------------------------------------------------
\begin{figure}[t]
\includegraphics[width=1.6in]{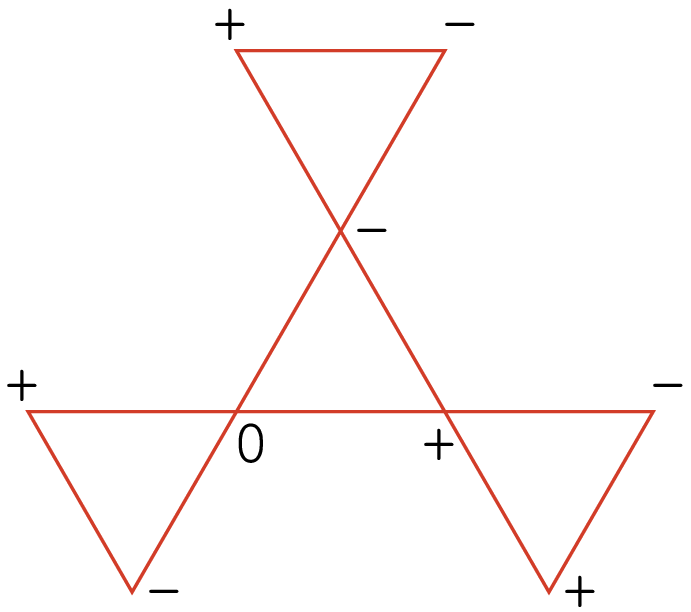}
\caption{ (color online).
A sample configuration in which the central triangle can resonate without
leaving the classical ground state manifold.
}
\label{fig:flippble}
\end{figure}
%---------------------------------------------------------------------------

The appearance of the VBS-T or VBS-B states can be qualitatively explained as follows.
Let us first consider the $D=0$ case. The classical model ($J_\perp=0$)
has the ground state
constraint such that each triangle on the kagome lattice should have 
one of the spin arrangements
$\texttt{++-}$, $\texttt{+--}$, or $\texttt{+0-}$ (and permutations), where
the eigenvalues of the spin-1 $S^z$ operator are denoted by $\texttt{+}$,
$\texttt{-}$, and $\texttt{0}$. There are many different ways to arrange
classical spins on the lattice to fulfill the above constraint, leading to a
highly degenerate classical ground state manifold. The third group
of states ($\texttt{+0-}$ and permutations) is quite special in the quantum
case --- a kinetic term (or an XY term) acting on such a state leaves
the spins on a triangle (a flippable triangle) in the classical ground
state manifold; for example,
$S^+_1S^-_2|\texttt{0+-}\rangle=|\texttt{+0-}\rangle$.
In general, such a move may violate the classical ground state constraint
on the neighboring triangles. However, it does not violate the constraint
in the special case when the three neighboring triangles of a flippable
triangle have $\texttt{+}$ and $\texttt{-}$ spins as shown in
Fig.~\ref{fig:flippble}.
Thus the flippable triangles can resonate without leaving the classical
ground state manifold. 
This consideration naturally leads to the following wave function that has 
a kinetic energy gain of $J_\perp$ (over the classical ground states)
in first order (degenerate) perturbation theory
$$
  |\psi\rangle=|\texttt{0+-}\rangle+|\texttt{0-+}\rangle
     +|\texttt{+0-}\rangle+|\texttt{-0+}\rangle
     +|\texttt{+-0}\rangle+|\texttt{0-+}\rangle.
$$

On the other hand, two flippable triangles shown in Fig.~\ref{fig:flippble}
can resonate simultaneously giving rise to a resonating bow-tie.
It is expected that turning on an infinitesimal $J_\perp$ will favor
the spin configurations where either the number of independently 
resonating triangles or the total number of resonating triangles
is maximized. In the latter case, we expect that
resonating bow-ties are close-packed as shown in Fig.~\ref{fig:bchi:d0b}.
This arrangement of bow-ties violates the classical ground state constraint
and one needs to project out the components of the wave function that violate the
constraint. \cite{moessner:sondhi} It is not possible to determine which
state has lower energy based on the above analysis.

Therefore, quantum
fluctuations lift the macroscopic degeneracy of the classical ground states
and gives rise to a VBS phase via quantum order by disorder mechanism.
The VBS phase should survive at small but finite values of $D/J_z$ 
because of a finite excitation gap.
The state with maximal number of independently flippable triangles is
exactly the VBS-T state that is found in quantum Monte Carlo simulations,
{\it i.e.}~the configuration shown in Fig.~\ref{fig:bchi:d0}.
The spin configuration, where all the
flippable triangles resonate, is the VBS-B phase and is shown in
Fig.~\ref{fig:bchi:d0b}. The energies of those two states are very close.

As we have noted above, the bond order is very strong in
the VBS-T(B) phase. This may be explained by the fact that the VBS-T phase
can be chosen via resonating triangles in first order perturbation theory 
whereas, in most of other cases (including VBS-H), plaquette 
resonance is obtained in higher order perturbation theory. 
As a result, the VBS-T(B) phase may be more robust than other cases.

%---------------------------------------------------------------------------
\subsection{Superfluid-paramagnet phase transition}

%---------------------------------------------------------------------------
\begin{figure}[t]
\includegraphics[width=3.0in]{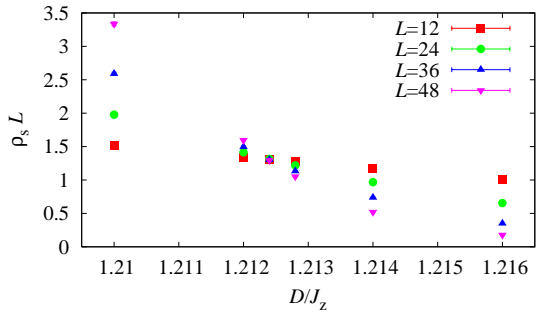}
\caption{ (color online).
Finite size scaling of the superfluid density $\rho_s$ at $J_z/J_\perp=6$
for $\beta/L=1/J_\perp$.
}
\label{fig:rhos:scaling}
\end{figure}
%---------------------------------------------------------------------------

%---------------------------------------------------------------------------
\begin{figure}[t]
\includegraphics[width=3.0in]{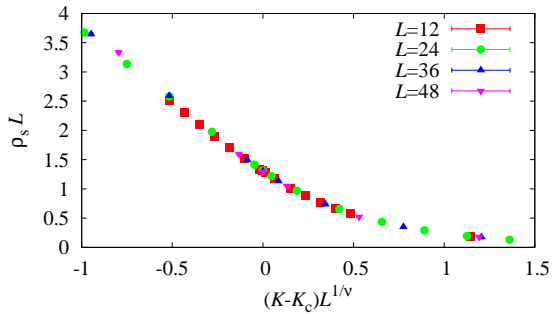}
\caption{ (color online).
Data collapse of the superfluid density $\rho_s$ at $J_z/J_\perp=6$ for
$\beta/L=1/J_\perp$.
}
\label{fig:rhos:collapse}
\end{figure}
%---------------------------------------------------------------------------

In this subsection, we discuss the phase transition between
the superfluid (or XY ferromagnet in spin language) 
and the featureless quantum paramagnet.
Our numerics shows that this transition is continuous and most likely
belongs to the 3d XY universality class. We analyze the finite size scaling
of our data as follows. We measure the superfluid density by measuring the
winding number fluctuations. \cite{windings} In the vicinity of a continuous
transition, the superfluid density scales as
\beq
  \rho_s=L^{-z} F_{\rho_s}(L^{1/\nu}(K_c-K), \beta/L^z),
  \label{rhos_scaling}
\eeq
where $F_{\rho_s}$ is the scaling function, $L$ is the linear system size,
$z$ is the dynamical critical exponent, $\nu$ is the correlation length
exponent, $K_c-K=(D/J_z)_c-D/J_z$ is the distance to the critical point,
and $\beta$ is the inverse temperature.
To cross the phase boundary, we change $D$ and keep $J_z$ fixed.
The data scale very well with the dynamical critical exponent $z=1$.
In Fig.~\ref{fig:rhos:scaling}, the superfluid density $\rho_s$ times
the system size $L$ is shown as a function of the coupling constant.
As follows from the above scaling form, the curves for different system
sizes should cross at the transition point if the ratio $\beta/L^z$ is
fixed. Such a distinct crossing point is seen at $(D/J_z)_c=1.2124$.
It also follows from Eq.~\ref{rhos_scaling} that the curves for
different system sizes should collapse onto a universal curve for
appropriate values of $\nu$ and $(D/J_z)_c$ when $\rho_sL$ is plotted as
a function of $[(D/J_z)_c-D/J_z]L^{1/\nu}$. In Fig.~\ref{fig:rhos:collapse},
we show such a data collapse for $\nu=0.67(2)$ and $(D/J_z)_c=1.2124(2)$.
The error bars are estimated from the stability of the data collapse with
respect to varying the fitting parameters. Thus one may conclude that the
superfluid-paramagnet quantum phase transition is continuous. One may also
infer that the transition is in the 3d XY universality class from $\nu=0.67(2)$.
However, we have not measured the other critical exponents that could be
used to unambiguously confirm this prediction.

%---------------------------------------------------------------------------
\subsection{Superfluid-VBS phase transitions}

%---------------------------------------------------------------------------
\begin{figure}[t]
\includegraphics[width=3.0in]{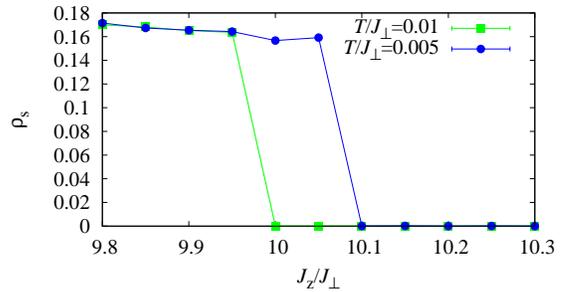}
\caption{ (color online).
The superfluid density $\rho_s$ as a function of $J_z/J_\perp$ for $L=12$,
$D=0$, and different temperatures. Lines are guides to the eye.
}
\label{fig:rhos:d0}
\end{figure}
%---------------------------------------------------------------------------

The quantum phase transitions out of the superfluid phase to VBS phases
are strongly first order. As an example, let us consider the transition from
the superfluid phase to the VBS-T phase.
In Fig.~\ref{fig:rhos:d0}, we show the superfluid density $\rho_s$ as
a function of $J_z/J_\perp$ at different temperatures. The superfluid density jumps
at large values of $J_z/J_\perp$ indicating a transition to an insulating
phase (VBS-T phase). We confirm that the transition is strongly first order
by observing hysteresis effects upon increasing or decreasing $J_z/J_\perp$
across the transition (not shown) and by a double peaked structure in the
distribution of the XY energy (kinetic energy of bosons),
$J_\perp \langle S^x_iS^x_j+S^y_iS^y_j\rangle$, close to the transition
(not shown).

The transition from the superfluid phase to the VBS-H phase is also first
order. In the top part of the VBS-H lobe (see Fig.~\ref{fig:phase:diagram}),
the first order nature becomes somewhat weaker 
as the value of $J_\perp/J_z$ is decreased.
We do not observe a direct transition from the VBS-H phase to the quantum
paramagnetic phase, which was predicted in Ref.~\onlinecite{xu:moore2}, in
the parameter region we can access, i.e. for $J_\perp/J_z \gtrsim 0.08$.
There is a narrow region of the superfluid phase between those two phases,
as shown in Fig.~\ref{fig:phase:diagram}. However, we cannot rule out
the possibility that there might be a direct
transition at much smaller values of $J_\perp/J_z$.

%---------------------------------------------------------------------------
\section{Discussion}

We now discuss possible origins of the discrepancy between
our quantum Monte Carlo results and those of previous analytic approaches with various 
approximation schemes. \cite{xu:moore1,xu:moore2,levin:wen}
In particular, it is found that the VBS-H phase, not the FP, is
the stable ground state for moderate strength of $D/J_z$ and small $J_{\perp}/J_z$.
This suggests that the string-net mean field theory analysis in a previous work \cite{levin:wen}
may not be sufficient for the identification of the true ground state. 

The string-net
picture starts from an alternative representation of the spins on the kagome lattice,
where the spins are considered to be in the middle of the links connecting the sites
of the honeycomb lattice (that can be obtained by connecting the centers of triangles
on the kagome lattice). \cite{levin:wen}
This honeycomb lattice is a bipartite lattice and
consists of $A$ and $B$ sublattices. The occupation of a given link $I = \langle ij \rangle$ by 
a string is defined as follows; the link contains an oriented string 
pointing from $i \in A$ to $j \in B$ if $S^z_I = +1$ and from $j$ to $i$ if
$S^z_I = -1$. The link is empty if $S^z_I = 0$. Then the spin configurations
on the original kagome lattice and those of oriented closed strings are in
exact correspondence. \cite{levin:wen}

The string-net mean-field theory uses the following ansatz for the
variational ground state wavefunction. \cite{levin:wen}
\beq
\Psi_{z} (X) = \prod_{ij} z_{ij}^{n_{ij}},
\eeq
where $X$ represents an oriented string configuration and $\{ z_{ij} \}$ 
a large number of variational parameters, and $n_{ij}$ the occupation 
number of the oriented link $ij$. The mean-field analysis begins with 
a (translationally-invariant) string liquid state (or a string condensed state) 
where $z_{ij}$ can be set to a constant $\alpha$. Considering the spectrum
of collective modes in this state and the identification of the wavevector where
the collective modes become soft, one can in principle study the instability 
to a translational-symmetry broken state. More explicitly, this can be 
achieved by writing $z_{ij} = \alpha e^{E_{ij} + i A_{ij}}$ and studying the fluctuations
of $E_{ij}$ and $A_{ij}$. It was found that the soft mode is described by
$E = \Phi E_{+{\bf Q}} + \Phi^*  E_{-{\bf Q}}$ and $A = 0$, where
$E_{\bf Q}$ is the Fourier mode at ${\bf Q} = (4\pi/3,0)$ and 
equivalent wavevectors. \cite{levin:wen}
Here $\Phi$ is a complex number.

The energy (or the Ginzburg-Landau theory)
of the system as a function of $\Phi$ can be obtained as \cite{xu:moore2,levin:wen}
\begin{equation}
H(\Phi) = A |\Phi|^2 + B |\Phi|^4 + C [\Phi^6 + (\Phi^*)^6] + \cdots, 
\end{equation}
where $A, B, C$ are real constants and $\Phi$ can be regarded as
an order parameter. 
The choice of the ground state
is sensitive to the phase of $\Phi$ and hence the sign of $C$ basically
determines the true ground state. It was found that if $C$ is positive (negative),
then the FP (VBS-H) is favored. \cite{levin:wen} This sign, however, is
very difficult to determine in analytic approaches. 
In Ref.~\onlinecite{levin:wen}, an unrestricted variational 
wavefunction calculation was also done on
a small system size $3 \times 3$, leading to the conclusion 
that the ground state may be the FP phase. \cite{levin:wen}
We think, however, that this system size is perhaps too small for
definitive conclusion. 
Indeed our quantum Monte Carlo results are clearly 
consistent with the VBS-H phase, and hence the negative 
sign of $C$ in the Ginzburg-Landau theory.

\acknowledgments

This work was supported by the NSERC, CRC, CIFAR,
KRF-2005-070-C00044, and the Swiss National Science Foundation (S. V. I.).
We thank Cenke Xu, Joel Moore, Michael Levin, Xiao-Gang Wen, and
Matthias Troyer for helpful discussions.

%---------------------------------------------------------------------------


\begin{thebibliography}{99}
\bibitem{remark}
On the time scale of our quantum Monte Carlo simulations, the system stays
in one of the six degenerate symmetry-broken states. However, the finite
size system should fluctuate between different degenerate states on much
longer Monte Carlo time scales.
\end{thebibliography}
\end{document}